\documentclass{article}

\usepackage[english]{babel}

\usepackage[letterpaper,top=2cm,bottom=2cm,left=3cm,right=3cm,marginparwidth=1.75cm]{geometry}

\usepackage{authblk}
\usepackage{amsmath}
\usepackage{graphicx}
\usepackage[colorlinks=true, allcolors=blue]{hyperref}
\usepackage{pgfplots}
\usepackage{adjustbox}

\title{Reinforcement Learning Jazz Improvisation: When Music Meets Game Theory}
\author[1,*,$\dagger$]{Vedant Tapiavala}
\author[1,$\dagger$]{Joshua Piesner}
\author[1]{Sourjyamoy Barman}
\author[1,2]{Feng Fu}

\affil[1]{Department of Mathematics, Dartmouth College, Hanover, NH 03755, USA}
\affil[2]{Department of Biomedical Data Science, Geisel School of Medicine at Dartmouth, Lebanon, NH 03756, USA}
\affil[*]{Correspondence: vedant.n.tapiavala.26@dartmouth.edu}
\affil[$\dagger$]{These authors contributed equally to this work.}
\begin{document}
\maketitle

\begin{abstract}
Live performances of music are always charming, with the unpredictability of improvisation due to the dynamic between musicians and interactions with the audience. Jazz improvisation is a particularly noteworthy example for further investigation from a theoretical perspective. Here, we introduce a novel mathematical game theory model for jazz improvisation, providing a framework for studying music theory and improvisational methodologies. We use computational modeling, mainly reinforcement learning,  to explore diverse stochastic improvisational strategies and their paired performance on improvisation. We find that the most effective strategy pair is a strategy that reacts to the most recent payoff (Stepwise Changes) with a reinforcement learning strategy limited to notes in the given chord (Chord-Following Reinforcement Learning). Conversely, a strategy that reacts to the partner's last note and attempts to harmonize with it (Harmony Prediction) strategy pair yields the lowest non-control payoff and highest standard deviation, indicating that picking notes based on immediate reactions to the partner player can yield inconsistent outcomes. On average, the Chord-Following Reinforcement Learning strategy demonstrates the highest mean payoff, while Harmony Prediction exhibits the lowest. Our work lays the foundation for promising applications beyond jazz: including the use of artificial intelligence (AI) models to extract data from audio clips to refine musical reward systems, and training machine learning (ML) models on existing jazz solos to further refine strategies within the game.

\vspace{0.5cm} 

\end{abstract}

\section{Introduction}    

The connections between mathematics and music are far-reaching~\cite{wright2009mathematics}. Much of music theory has been devoted to understanding musical perception through the lens of mathematics. Music can be described in the language of ratios, geometric objects, Fourier analysis, and much more. However, the connections between music and game theory have largely been unexplored. In this paper, we will bridge this connection by using game theory to model duo jazz improvisation through the lens of reinforcement learning.

Game theory is a natural choice to describe musical improvisation~\cite{goldman2020improvisation}. A mathematical game involves two or more players who choose from a set of options with limited information and then incur a corresponding payoff based on which option they and the other player(s) chose, leading to uncertainty~\cite{fudenberg1991game}. In group musical improvisation, a given player (musician) has an array of options (the notes available) and gets to decide which of these options to play. That player then receives a resulting payoff: ``Within the context of what everybody else played, did what I just play sound good or bad?'' The mission of a rational player (as is the case in all mathematical games) is to maximize their payoff (in other words play an improvisation that sounds as good as possible).

Few prior work has systematically explored this idea. Leslie and Hassanpour~\cite{originalPaper} have introduced the idea that simple improvisation can be modeled using a mathematical game in which each player gets to pick between two possible notes. The simplest version of this game would be players each picking random notes to create music. Developing upon this, Leslie and Hassanpour find that strategies such as reinforcement learning tend to converge to repeatedly playing the notes that are most harmonious. Several composers have also explored the idea of stochastic music. Hiller and Isaacson~\cite{sandred2009revisiting} were the first with their Illiac Suite in 1957 which employed various probabilistic algorithms to compose each movement of the suite. Later, Iannis Xenakis expanded upon this idea with compositions such as \textit{Metastaseis}, \textit{Pithoprakta}, and \textit{Achorripsis}~\cite{serra1993stochastic, arsenault2002iannis}.  In the past 50 years, a select handful of inventors, music theorists, and computer scientists, such as Johnson-Laird, Pressing, and Gifford, have tried to create machine algorithms that improvise music~\cite{johnson2002jazz, pressing1984cognitive, gifford2018computational}. Attempts have ranged from using Markov models to machine learning~\cite{doornbusch2010algorithmic, chan2006improving,huang2024symbolic}. While none of these composers and music theorists considered the use of game theory, they developed possible algorithmic strategies that could have been used in a musical game. In our paper, we will test musical strategies similar to the algorithms used by Hiller, Isaacson, Xenakis, and those mentioned by Gifford~\cite{sandred2009revisiting,serra1993stochastic,gifford2018computational}, but more precisely within a game-theoretical setting. Additionally, we will test \emph{reinforcement learning} strategies, which are widely used in game theory models in economics~\cite{chang2021multi,huo2017risk}, among other fields, such as in market failure~\cite{mesly2023role, fudenberg2016whither}. To evaluate these strategies, we will employ an expanded version of the game model proposed by Leslie and Hassanpour~\cite{originalPaper}.

Additionally, while the exploration of stochastic music and the application of game theory to improvisation have largely taken place in the context of Western Classical Music, our paper will focus on Jazz Music.

Our reason for focusing on jazz is simple. Improvisation is at the heart of jazz. No other genre of music uses improvisation as much~\cite{goldman2020improvisation}. In a typical jazz song, musicians will constantly improvise at varying levels of freedom throughout. The improvisation reaches its highest intensity in the so-called ``solo section'' ~\cite{monson2009saying}. This is a section in nearly every jazz song, where one or more players are given the opportunity to freely improvise for an extended amount of time over an agreed upon set of chords. This set of chords is known as as a chord progression.

\subsection{Relevant Music Theory}

Harmony has been historically measured on a scale from consonance to dissonance, with consonant harmony typically sounding more harmonious and dissonant harmony sounding less harmonious~\cite{cazden1980definition}. Nearly all the information of whether two notes sound harmonious or not comes from the base frequencies (also called fundamental frequencies) of each note. Therefore, there have been several attempts to find mathematical rules that determine whether two frequencies create consonant or dissonant harmony. The leading idea is that consonance is related to the ratio between two frequencies~\cite{kameoka1969consonance}. Specifically, if the ratio can be simplified to a fraction that is relatively ``simple,'' or in other words the numerator and denominator are small, then those two frequencies are said to be consonant. For example, if the two frequencies were 440 and 220 the ratio between the two would simplify to 2:1 which is very simple. (These two frequencies actually correspond to two A's of octave apart). If the two frequencies were 440 and 501, this would simplify to 440:501, which is not simple. This would correspond to a dissonant frequency. This is the approach used in this paper.

It should be acknowledged that there are several other theories that attempt to mathematically explain harmony. Some of these include looking at relative periodicity, the harmonic series, and overlapping of overtones~\cite{stolzenburg2015harmony}. Most recently, Ref.~\cite{gonzalez2023quantifying}
finds an interesting ``first-decline-and-then-increase'' pattern in the evolution of harmony by using quantitative measures of key uncertainty, diversity and novelty over the past 400 years of Western classical music.

\section{Methodology and Model}

The most ubiquitous chord progression in jazz is perhaps the ``blues'' chord progression. There are many variations of this chord progression, but in our case we will focus on the Bb Quick Change Blues Chord Progression~\cite{broze2013diachronic}. Specifically, we will do a 12-bar Blues: Bb7, Eb7, Bb7, Bb7, Eb7, Eb7, Bb7, Bb7, Cm7, F7, Bb7, F7. The 12-bar Blues was chosen since it is the most common length for a blues~\cite{meadows1991improvising}. However, we believe the results would be equivalent for any variation chosen.

Because we have chosen the 12-bar blues, this means that the chord progression will cycle every 12 measures. The blues are also typically in the time signature ${}^{4}_{4}$. If we subdivide our measure into eighth notes that leaves us with 8 beats per bar. Multiplied by the 12 bars in a cycle, we get 96 beats per cycle. A typical jazz solo usually lasts for multiple cycles. While no particular number was found for an ideal number of cycles for a jazz improvisation game, many forms of music often utilize phrases that last eight cycles~\cite{hughes2011}. Thus, the length of 8 cycles for a given game was selected, which leaves us with 768 beats. 

Our model of the jazz improvisation game includes two players each choosing integer sound frequencies between 28 and 4186 Hz, the lowest and highest frequency notes available on a standard piano. Both players then simultaneously play their notes together without knowledge of what the other player is going to play. This repeats for every beat in the song, with each player able to see the history of all notes played before but unable to know what the other player will exactly play next. This results in a kind of improvisatory duet. In addition, to account for the chord progression, there are 4 extra notes that are played every beat which correspond to the notes within the chord for that measure (every 8 beats is a new measure). These 4 extra notes are outside of the control of both players and are simply part of the game. However, both players have pre-existing knowledge on what these ``chord notes" will be. In our simulation, players had one of nine strategies (five non-reinforcement learning strategies described in Table 1 and four reinforcement learning strategies described in Table 2), and payoffs were calculated based on variance and harmony scores. These scores measured how varied a player's note choices were and how harmonious those notes were with the current chord progression and the other player's selected note.

\subsection{Payoff Calculation}
The payoff was based on two separate score calculations: the variance $V$ and harmony $H$ scores.

\subsubsection{Variance Score (Diversity)}
Variance was used to prevent players from playing the same notes repetitively. Instead, different notes are encouraged to play because the music would be boring otherwise. The idea of variance in generative music has been explored previously~\cite{berndt2012}, but few (none to the knowledge of the authors) mathematical models have been proposed. While statistical variance was initially used as a proxy calculation for musical variance, statistical variance incentivizes the same notes being played repetitively to increase the difference in counts of notes. For example, if A has already been played 10 times and no other notes have been played, playing A again will increase the statistical variance (11 with the remaining list being filled with zeros).

Instead, Shannon's Diversity Index, specifically through species evenness, was used as a proxy calculation for musical variance~\cite{pielou1966measurement}. Species evenness typically measures how evenly species are divided within a population. In this paper, we utilize this diversity measure as a proxy to see how evenly notes are divided within the music. The notes were thus represented analogously as different species, with the number of species used in the species evenness calculations being analogous to the number of each note that has been played by the time of the variance score calculation. The variance score is calculated every beat (since the payoff is calculated every beat).

Let $p_i$ represent the percent of how many times note $i$ has been played. Let $n$ represent the number of notes played in the simulation thus far. There are 12 potential notes that can be played in the Western tuning system (different octaves are treated as the same note). Our calculation for the variance score is represented in Equation 1.

\begin{equation}
    V=\frac{{\sum_{i=1}^{n} \left( -\frac{p_i}{\ln(p_i)} \right)}}{\ln(n)},\quad n=12.
\end{equation}

Under this accounting, a piece that plays each note an equivalent number of times would achieve the highest variance score. However, this extreme edge case is likely not ideal for music and could be refined in future research.

\subsubsection{Harmony Score}
The harmony score looks at six notes, one played by each of the two players and the four representing the current chord in the chord progression. Each possible pair of these six notes is put into a fraction, which is then simplified. Then, the numerator and denominator are added to one another. These numerator/denominator sums are added and averaged to get the harmony score, where a smaller harmony score is better. Let \( V = \{v_1, v_2, \ldots, v_n\} \) be the set representing a chord, where \( n = 6\) is the number of notes in the chord and \( v_i \) is the frequency of a given note in the chord. Then, the harmony score \( H \) of the chord can be calculated as follows:
\begin{equation}
H = \frac{1}{\binom{n}{2}} \sum_{\{v_i, v_j\} \subseteq V, i \neq j} \left(\frac{v_i + v_j}{\gcd(v_i, v_j)}\right), \quad n = 6,
\end{equation}
where $\gcd$ defines a function to find the greatest common denominator of its parameters. Equation 2 can be rewritten as follows for simplicity:
\[
    H=\frac{\sum_{i=1}^{n-1} \sum_{j=(i+1)}^{n} (\frac{\nu_i + \nu_j}{\gcd(\nu_i, \nu_j)})}{\frac{n(n-1)}{2}}, \quad n=6.
\]

\subsubsection{Multiplication Factor}
Since Shannon's Diversity Index, utilized in the variance score calculation, ranges from 0 to 1, the variance score was continually lower than the harmony score, which adds a numerator and denominator. Thus, the harmony score was guaranteed to be at least two and was usually around 1200. To account for this, the variance score was multiplied by a multiplication factor to ensure that the variance and harmony scores were comparable and that there was no immediate benefit of trade-offs of variance for harmony, or vice versa. 

The multiplication factor was calculated by simulating ten jazz improvisation games of two random players. For each simulation, both the variance and harmony scores were averaged. Since the variance scores for improvisation games of two purely random players should approach 1 as the game continues, the average of the harmony scores was deemed equivalent to the multiplication factor. In other words, the multiplication factor, if applied in that trial, would have made the payoff negligible. This goal of negligible payoff for Random-Random jazz improvisation game simulations was made due to the random strategy's role in this game as a control.

Each trial's ideal multiplication factor was then averaged and used as a constant for the rest of the simulation. This final multiplication factor was found to be  $\sim 1208.7571$ in our simulations, represented as $M$. Thus, each variance score was multiplied by $M$ before being used in the payoff calculation to create a $50-50$ weighting of the importance of variance and harmony.

After calculating the variance score, denoted $V$, and calculating the harmony score, denoted $H$, the payoff was calculated through the formula represented in Equation 2. This formula accounts for a lower harmony score being better, while a higher variance score is better, according to this paper's model.

\begin{equation}
    P = \frac{VM-H}{VM+H}
\end{equation}

Since the variance score relies on past notes, the variance score, after being multiplied by the multiplication factor, for the first beat was artificially set to equal the harmony score. Thus, the payoff of the first beat will always equal 0.

For the final results, $100$ trials of each possible pair of strategies was run. For example, the randomness strategy was played against all eight other strategies and itself $100$ times. Then, the $100$ trials' payoffs were averaged for a resulting payoff. Each trial included $96$ measures of eight beats each, with each beat representing a round. In this round, each player played one note. Thus, each trial involved $768$ beats, with two player-chosen notes each. Over 100 trials, over $150,000$ notes were played for each pair, resulting in almost $7,000,000$ notes being played over the course of the simulation.

\subsection{Improvisation Strategies}
In total, nine strategies were tested: Randomness, Chord Following, Scale Following, Harmony Prediction, Stepwise Changes, Simple Reinforcement Learning, Chord-Following Reinforcement Learning, Chord-Specific Reinforcement Learning, and Two-Player Reinforcement Learning. These strategies are split into two categories: non-reinforcement learning (Table 1) and reinforcement learning (Table 2) strategies. This categorization is based on a sole criterion: whether a strategy utilizes data about prior payoffs, necessary for the trial-and-error character of reinforcement learning algorithms~\cite{kaelbling1996reinforcement}.

\begin{table}[h!]
    \centering
    \caption{Non-Reinforcement Learning Player Strategies in the Game}
    \adjustbox{width=\textwidth}{%
    \begin{tabular}{|p{2cm}|p{13.5cm}|}
        \hline
        \textbf{Strategy} & \textbf{Description} \\
        \hline
        Randomness Strategy & Used as a control, from which to measure the effectiveness of other strategies. A player using this strategy will select a note, at random, from between 28 and 4186 Hz. \\
        \hline
        Chord Following Strategy & Players follow the chord progression, meaning they only play notes, or octaves of notes, in the chord currently being played. The Bb chord progression has twelve chords, each voiced with four notes. The octaves of a note X can be determined by multiplying or dividing the frequency corresponding to X by whole numbers, given the new frequencies remain within the game's defined frequency range. \\
        \hline
        Scale Following Strategy & Players follow the scale and only play notes, or octaves of notes in the Bb mixolydian scale. This scale following provides 51 different notes for the player to randomly select from. This strategy is inspired by a common beginner-to-intermediate method of jazz improvisation involving choosing a scale to improvise over that fits with notes in the chord progression, with the mixolydian scale being a common choice ~\cite{meadows1991improvising}. \\
        \hline
        Harmony Prediction Strategy & Players play a note that would harmonize well with the last note played by the opposing player. This strategy determines a harmonious note by multiplying or dividing the frequency corresponding to the note by whole numbers, provided that these frequencies remain within the defined frequency range. If the last note played is a prime number, 1 Hz would be added to make it divisible by other whole numbers. \\
        \hline
    \end{tabular}%
    }
\end{table}
\begin{table}[h!]
    \centering
    \caption{Reinforcement Learning Player Strategies in the Game}
    \adjustbox{width=\textwidth}{%
    \begin{tabular}{|p{2cm}|p{13.5cm}|}
        \hline
        \textbf{Strategy} & \textbf{Description} \\
        \hline
        Simple Reinforcement Learning & Determines a set of weights, one corresponding to each note, similar to the Roth-Erev reinforcement learning approach ~\cite{roth1995learning}. After each beat, the player adds the resulting payoff to the weight of the note that it just played updating the probability for each note to be played. Thus, a negative payoff reduces the likelihood that the previous note will be played again, while a positive payoff increases this likelihood. \\
        \hline
        Chord-Following Reinforcement Learning & Derived from the chord-following strategy. Thus, players following this strategy will only play notes in the current chord, or octaves of those notes. However, this strategy adds to the chord-following strategy by having a set of weights for each possible note. Similar to the simple reinforcement learning strategy, payoffs from each beat update the weights by adding the payoff of the last beat to the weight of the note that was just played. \\
        \hline
        Chord-Specific Reinforcement Learning & Employs four isolated simple reinforcement learning strategies, one for each type of chord in the chord progression. In other words, the player uses a different set of weights depending on the current chord (eg: Bb7) and will update only the set of weights corresponding to that current chord. \\
        \hline
        Two-Player Reinforcement Learning & Similar to the simple reinforcement learning strategy but with more data. Instead of adding the most recent payoff to the weight of the note that was just played, the payoff is added to the weights for both notes the players just played. Thus, this strategy uses the simple reinforcement learning strategy but with twice the amount of data. \\
        \hline
        Stepwise Change Strategy & A form of reinforcement learning. If the previous payoff was positive, the player will ``hop'' by a multiple of the note (either a fraction of 1 over a whole number or multiplying it by a whole number). For example, a player who previously played a 400 Hz note with a positive payoff has the same likelihood of hopping to a 200 Hz note as an 800 Hz note.
        If the previous payoff was negative, the player will ``step'' by a multiple of the note (multiply or divide by a number between 1 and 1.1), creating a subtle tonal change. \\
        \hline
    \end{tabular}%
    }
\end{table}
\section{Results}

To analyze our jazz improvisation model, we utilized computer simulation in a Java program, the GitHub repository of which is provided in the end of this paper.

\begin{figure}[htb]
  \centering
  \includegraphics[width=0.8\textwidth]{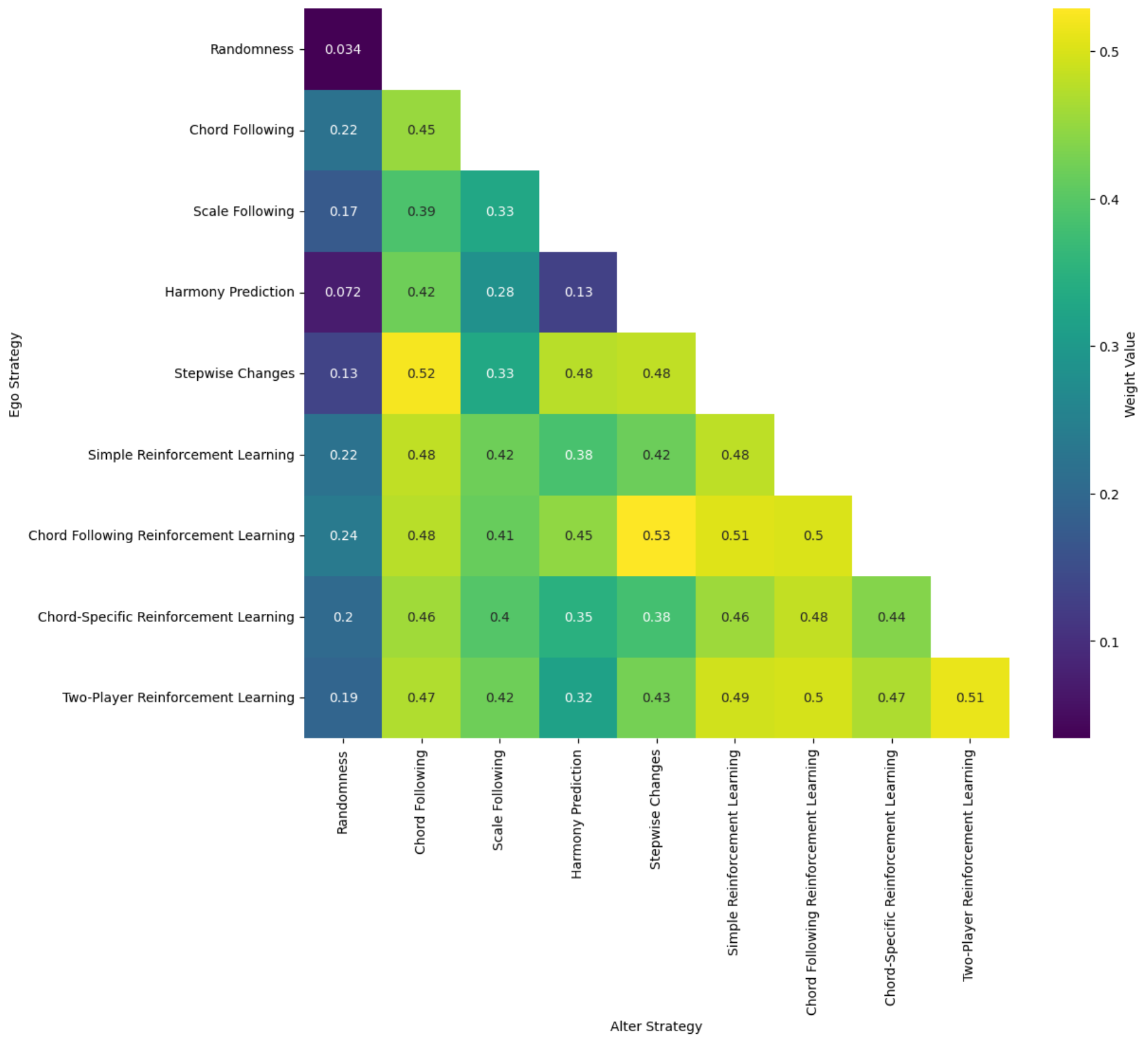}
  \caption{Heatmap Indicating Strategy-Strategy Payoffs. \newline R: Randomness, CF: Chord Following, SF: Scale Following, HP: Harmony Prediction, SC: Stepwise Changes, SRL: Simple Reinforcement Learning, CFRL: Chord-Following Reinforcement Learning, CSR: Chord-Specific Reinforcement Learning, TPRL: Two-Player Reinforcement Learning}
  \label{fig1}
\end{figure}

The average payoff of each strategy pair after $100$ trials was computed and is displayed to five decimal places in Table 3, also visualized as a heatmap in Figure~\ref{fig1}. The average of each strategy's payoff was computed and displayed in the bottom row of Table 3, also as a bar chart in Figure~\ref{fig:bar_graph} . The table also includes ``-" for any strategy that was played in a different part of the table (i.e., one player playing a Randomness strategy and another playing a Chord Following strategy was shown in the Randomness column, while a ``-" was shown in the Chord-Following column since it was already shown). Thus, the row denoting the average payoff of each strategy does not equal the average of the column (other than for the Randomness strategy) since it accounts for every time the strategy was played. For example, for the Two-Player Reinforcement Learning strategy, the average payoff included the entire Two-Player Reinforcement Learning row and the column (but not double-counting two players playing this strategy at the same time).

\begin{table}[h]
\centering
\begin{tabular}{l|lllllllll}
\hline
&R & CF & SF & HP & SC & SRL & CFRL & CSR & TPRL \\ \hline
R & 0.03384 & - & - & - & - & - & - & - & - \\
CF & 0.21966 & 0.45162 & - & - & - & - & - & - & - \\
SF & 0.17260 & 0.38764 & 0.32780 & - & - & - & - & - & - \\
HP & 0.07237 & 0.41900 & 0.28182 & 0.12942 & - & - & - & - & - \\
SC & 0.13404 & 0.52125 & 0.33031 & 0.47558 & 0.48120 & - & - & - & - \\
SRL & 0.21967 & 0.48145 & 0.41864 & 0.38419 & 0.41681 & 0.47916 & - & - & - \\
CFRL & 0.23584 & 0.47615 & 0.41386 & 0.44727 & 0.52872 & 0.50505 & 0.50015 & - & - \\
CSR & 0.20317 & 0.45845 & 0.39573 & 0.34865 & 0.38134 & 0.45682 & 0.48276 & 0.43601 & - \\
TPRL & 0.19336 & 0.46914 & 0.41905 & 0.31650 & 0.42581 & 0.49365 & 0.49746 & 0.46694 & 0.51225 \\
\hline
\textbf{Average} & 0.16495 & 0.43160 & 0.34972 & 0.31942 & 0.41056 & 0.42838 & 0.45414 & 0.40332 & 0.42157 \\ \hline
\end{tabular}
\caption{Mean Payoff of Each Strategy Pair. \newline R: Randomness, CF: Chord Following, SF: Scale Following, HP: Harmony Prediction, SC: Stepwise Changes, SRL: Simple Reinforcement Learning, CFRL: Chord-Following Reinforcement Learning, CSR: Chord-Specific Reinforcement Learning, TPRL: Two-Player Reinforcement Learning}
\label{tablemp}
\end{table}

Upon an individual examination of each strategy pair, the strategy pair with the highest payoff was one player using a Chord Following Reinforcement Learning strategy and the other applying a Stepwise Changes strategy. The implications of this are further elaborated upon in the Discussion section.

On the other hand, the strategy pair with the lowest payoff were two Random players, which makes sense given the model's selection of two players playing Randomness strategies as the control. While the average payoff of this strategy over 100 trials was not exactly zero, this payoff was the closest to zero. The lowest non-control strategy pair average payoff was that between a player playing a Randomness strategy and a Harmony Prediction strategy. A low payoff from this strategy pair is expected since the harmony prediction relies on knowledge of the last note the other player played and the assumption that the next note will be similar, an invalid assumption when playing with a player playing a Randomness strategy.

The lowest payoff of a strategy pair not involving any randomness and thus a more realistic strategy was that of two players playing Harmony Prediction strategies at 0.12942.

The average payoff of each strategy is also shown in Table~\ref{tablemp}. This measure is a good indication of what strategy generally does well, regardless of the other player's strategy. While some strategies may do well in tandem with other specific strategies, selected strategies with high mean payoffs will likely do well regardless.

The strategy with the highest mean payoff was the Chord-Following Reinforcement Learning strategy, while the strategy with the lowest mean payoff was the Randomness strategy. However, since Randomness was used as a control in this simulation, the realistic strategy with the lowest mean payoff was the Harmony Prediction strategy. This strategy varied widely with its payoffs, performing well with high performing strategies and poorly with low performing strategies. Paired with the Stepwise Changes and Chord Following Reinforcement Learning strategies (the two with the highest mean payoffs), it received payoffs of 0.47558 and 0.44727 respectively. Paired with Random and itself (the two lowest mean payoffs), it received low payoffs of 0.07237 and 0.12942 respectively.

\begin{figure}
    \centering
    \begin{tikzpicture}
    \begin{axis}[
        xbar, 
        bar width=0.5cm,
        width=0.9\textwidth,
        height=10cm, 
        xlabel={Average Payoff},
        ylabel={Strategies},
        ytick=data,
        yticklabels={
            R,
            CF,
            SF,
            HP,
            SC,
            SRL,
            CFRL,
            CSR,
            TPRL
        },
        y dir = reverse,
        y tick label style={rotate=0, anchor=east},
        y label style={at={(axis description cs:-0.1,0.5)}, anchor=south},
        xmin=0,
        xmax=0.55,
        xticklabel style={
            /pgf/number format/fixed,
            /pgf/number format/precision=5
        },
        scaled ticks=false, 
        tick label style={font=\small} 
        ]
        \addplot[fill=gray] coordinates {
            (0.17428,1)
            (0.44643,2)
            (0.36370,3)
            (0.38393,4)
            (0.47398,5)
            (0.49295,6)
            (0.49906,7)
            (0.47568,8)
            (0.45550,9)
        };
    \end{axis}
    \end{tikzpicture}
    \caption{Average Payoff by Strategy.
    \newline R: Randomness, CF: Chord Following, SF: Scale Following, HP: Harmony Prediction, SC: Stepwise Changes, SRL: Simple Reinforcement Learning, CFRL: Chord-Following Reinforcement Learning, CSR: Chord-Specific Reinforcement Learning, TPRL: Two-Player Reinforcement Learning}
    \label{fig:bar_graph}
\end{figure}

 Of particular interest as well were the standard deviations of the trials done per pair. These were calculated by computing the standard deviation of the average payoffs of each trial. Some trials had very significant standard deviations, indicating that their payoffs varied a lot and were not as reliable as those strategies with lower standard deviations. All standard deviations of each strategy pair's 100 trials are shown in Table 4 to five decimal places.

 \begin{table}[!ht]
    \begin{center}
    \resizebox{\textwidth}{!}{%
    \begin{tabular}{l|ccccccccc}
    \hline
    & R & CF & SF & HP & SC & SRL & CFRL & CSR & TPRL \\ 
    \hline
    R & 0.00838 & - & - & - & - & - & - & - & - \\
    CF & 0.00875 & 0.00733 & - & - & - & - & - & - & - \\
    SF & 0.00906 & 0.00978 & 0.01002 & - & - & - & - & - & - \\
    HP & 0.00980 & 0.00902 & 0.01304 & 0.26052 & - & - & - & - & - \\
    SC & 0.01276 & 0.04401 & 0.01847 & 0.11258 & 0.10023 & - & - & - & - \\
    SRL & 0.00973 & 0.00989 & 0.01065 & 0.01758 & 0.03088 & 0.01251 & - & - & - \\
    CFRL & 0.00837 & 0.00779 & 0.01050 & 0.01465 & 0.05715 & 0.01130 & 0.00859 & - & - \\
    CSR & 0.00900 & 0.00924 & 0.01047 & 0.01269 & 0.02483 & 0.01026 & 0.01008 & 0.01053 & - \\
    TPRL & 0.00840 & 0.01011 & 0.00959 & 0.01468 & 0.05162 & 0.01259 & 0.01393 & 0.01310 & 0.01306 \\
    \hline
    \textbf{Average} & 0.00936 & 0.01288 & 0.01129 & 0.05162 & 0.05028 & 0.01393 & 0.01582 & 0.01224 & 0.01634 \\
    \hline
    \end{tabular}
    }
    \end{center}
    \caption{Standard Deviations of Payoffs by Strategy Pair. \newline R: Randomness, CF: Chord Following, SF: Scale Following, HP: Harmony Prediction, SC: Stepwise Changes, SRL: Simple Reinforcement Learning, CFRL: Chord-Following Reinforcement Learning, CSR: Chord-Specific Reinforcement Learning, TPRL: Two-Player Reinforcement Learning}
    \label{tablesd}
\end{table}

  \begin{figure}[htb]
  \centering
  \includegraphics[width=1\textwidth]{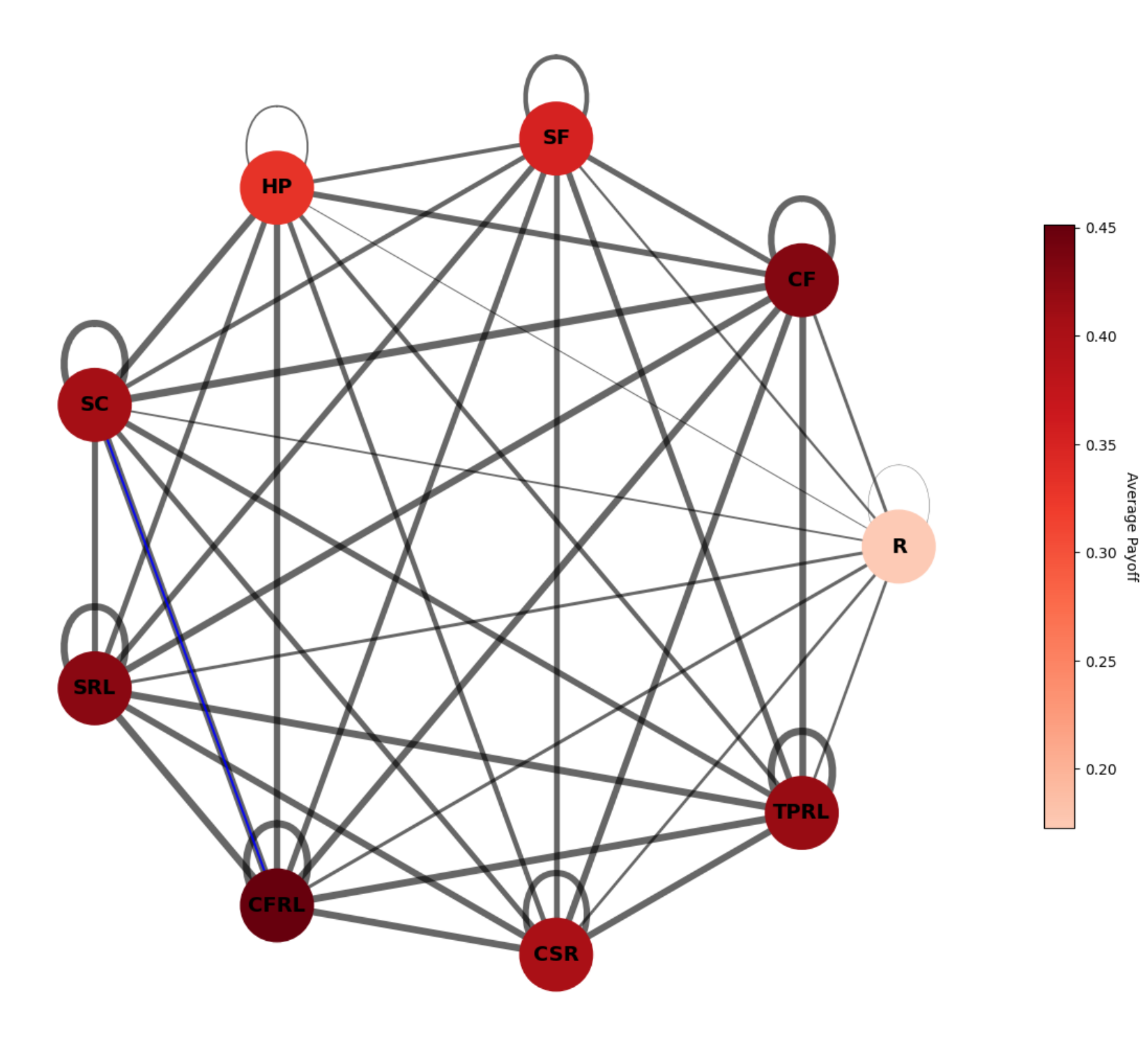}
  \caption{Network Graph Representing Payoff Relationships Between Strategies. \newline R: Randomness, CF: Chord Following, SF: Scale Following, HP: Harmony Prediction, SC: Stepwise Changes, SRL: Simple Reinforcement Learning, CFRL: Chord-Following Reinforcement Learning, CSR: Chord-Specific Reinforcement Learning, TPRL: Two-Player Reinforcement Learning}
  \label{fig3}
\end{figure}

As shown in Figure~\ref{fig3}, the strategy pair with the lowest standard deviation after 100 trials (the most consistent strategy) involved two players playing Chord Following strategies, indicating that the Chord Following strategy playing against itself produced highly consistent results compared to other strategy pairs. On the other hand, the most inconsistent strategy pair (that with the highest standard deviation) were two players playing Harmony Prediction strategies.

 Another point of note was the varying performance of reinforcement strategies. Each reinforcement strategy had slightly different performances, but each of the strategies did extremely well compared to most non-reinforcement strategies. The chord-following reinforcement strategy was the best strategy from the simulation in terms of average payoff, while the worst reinforcement strategy, the chord-specific reinforcement learning, was still better than three other strategies.

\begin{figure}[htbp]
  \centering
  \includegraphics[width=1\textwidth]{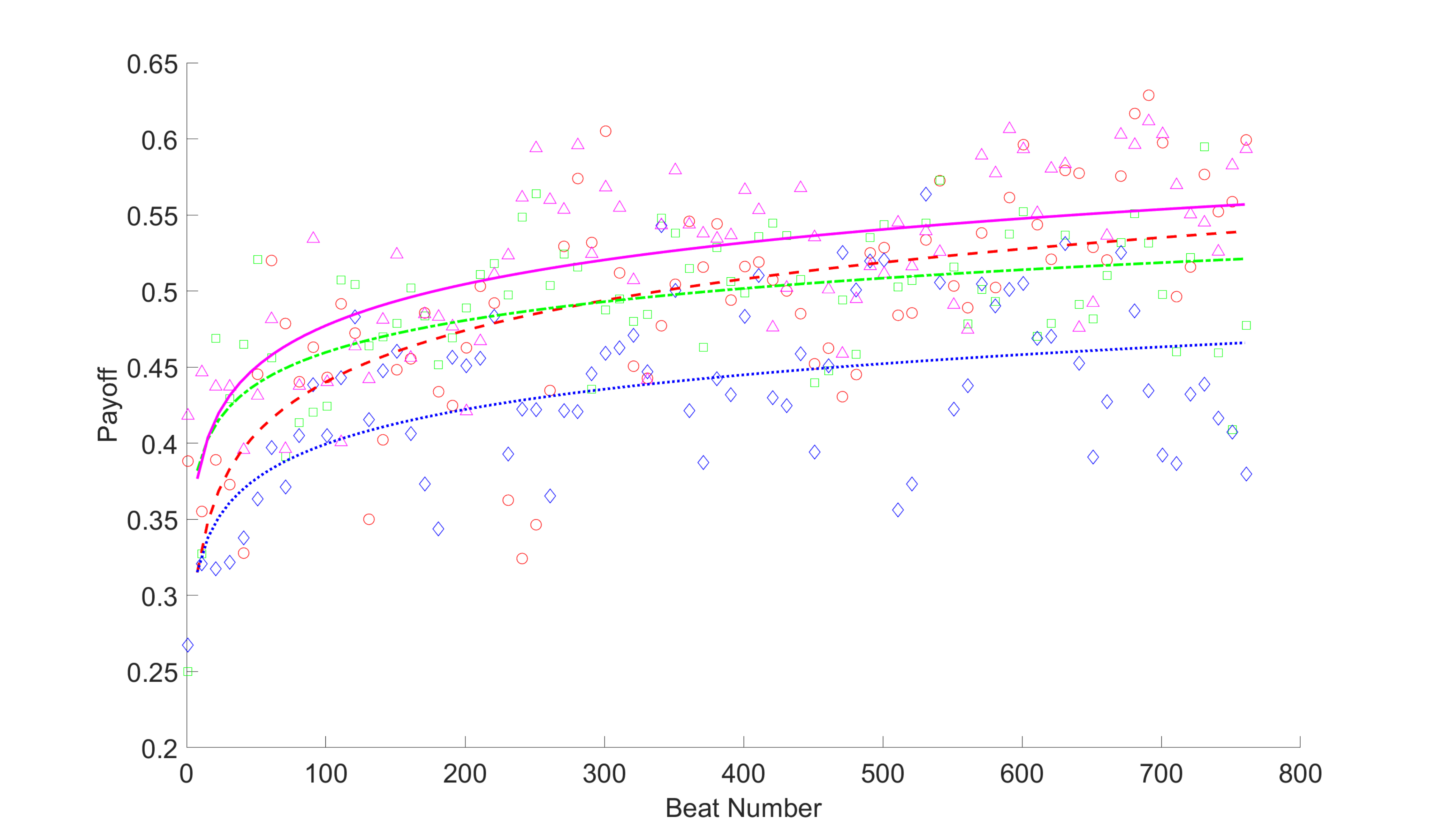}
  \caption{Reinforcement Learning Payoffs over Time. \newline Red Circles and Red Dashed Line: Simple Reinforcement Learning, Green Squares and Green Dash-Dot Line: Chord-Following Reinforcement Learning, Blue Diamonds and Blue Dotted Line: Chord-Specific Reinforcement Learning, Pink Triangles and Pink Solid Line: Two-Player Reinforcement Learning}
  \label{fig4}
\end{figure}

 The most important point to note from the reinforcement learning strategies, however, was the learning algorithms and how the payoffs changed over time. For this, a representative trial was taken and the payoffs were graphed for each reinforcement learning strategy pair (a strategy playing against itself) in Figure~\ref{fig4}. Since there were 768 points for each reinforcement learning, every tenth point was plotted. Furthermore, to counteract variability in payoffs due to inherent randomness, a moving window comprising 3\% of the game was utilized (23 point window size).
 A logarithmic regression using all 768 points per strategy was also adopted to demonstrate the reinforcement learning strategy's payoff changes over time. A logarithmic regression was used over other types of regression models since the R and R\textsuperscript{2} values were highest for this form of regression. The R values when using a log regression model for the reinforcement learning strategies were 0.44103, 0.37904, 0.32294, and 0.48744 for simple reinforcement learning, chord-following reinforcement learning, chord-specific reinforcement learning, and two-player reinforcement learning, respectively. Thus, all correlations between predicted values based on the logarithmic regression and the actual values were generally moderately positively correlated.

\section{Discussion}

Reinforcement Learning and Stepwise Change strategies performing well indicates that musicians in the game should be open to switching based on new feedback. In practical real-life terms, this would correspond to a musician shifting strategies after hearing a few notes clash with the rest of the music. 

The best strategy pair being the Chord Following Reinforcement Learning and Stepwise Changes strategy is also noteworthy. These two strategies are both forms of learning strategies, where the player learns more about the game through some new information. The Chord Following Reinforcement Learning player updates probabilities of playing certain notes based on previous payoffs. The Stepwise Change player looks at the previous payoff and its own previous note. If the payoff was good, the Stepwise Changes player will ``step,'' meaning it will slightly change the note. This note is still likely to be harmonious with the new note from the Chord Following Reinforcement Learning player.

On the other hand, the Harmony Prediction strategy with itself produced the lowest average payoff of any strategy excluding pairs where one or more was using the Randomness strategy. This strategy pair also had the highest standard deviation. This means it had some high payoffs and some extremely low payoffs. This is partly because the Harmony Prediction strategy always reacts to the last note played by the opponent strategy. Since both players are reacting to the other, it is possible for the players to get stuck in a loop in which similar notes are played over and over again, which would lead to a low variance score.

Furthermore, the good performance of Chord-Following strategies suggests that musicians should prioritize playing music that would harmonize well with the current chord. Since the Chord-Following reinforcement learning strategy had the highest average payoff out of all tested strategies, having different notes to play based on the current chord is a viable strategy for ``better'' (as defined by our simplified model) music. Additionally, the chord following strategy playing against itself without reinforcement learning had the second highest average payoff and was the most consistent strategy pair in the simulation, defined as having the lowest standard deviation amongst its 100 trials. This data indicates that these Chord-Following strategies are valuable for getting a more ``guaranteed'' way of having a good result in the model. This qualitatively matches real experiences in jazz improvisation. Often, as an improvisation exercise, beginners will be instructed to try playing only notes within the chord, as it is understood this strategy has very low risk.

With this in mind, it is surprising that Scale Following strategy had relatively low payoffs, given that scale following is a common strategy employed by actual musicians. Clearly, the strategy of randomly playing notes in a scale is not good enough to generate a high payoff. One possible explanation is that the extensive range of options available in scales increases the likelihood of playing dissonant intervals. Thus, if you pick two random notes from the Bb mixolydian scale, they will likely not harmonize with each other or with the given chord. 

Interestingly, the Scale Following strategy's performance fell into two distinct categories: low (relative to other payoffs in the simulation) non-reinforcement learning pairs and average reinforcement learning pairs. In other words, this strategy performed best when the opposing player had knowledge of the Scale Following player's past notes. This makes sense because the Scale Following player is limited to a very strict set of notes, so it is easier for a reinforcement strategy to predict what the Scale Following strategy will play next. This is analogous to how a soloist who plays only notes in a certain scale can be easier to accompany because the tonality of that player is predictable.

When observing reinforcement learning algorithms, all payoffs for these algorithms tended to increase time (see Figure~\ref{fig4}). The strongest correlation with predicted values in a logarithmic regression was seen in two-player and simple reinforcement learning strategies. The first of these two strategies, two-player reinforcement learning, had significant data advantages since it utilized data from both jazz improvisers. These data advantages are the likely cause for this model's strong performance increase over time. Both the simple and two-player reinforcement learning strategies also did not seem to plateau by the end of the game, indicating possible additional improvement if the game were to be continued for longer. A logarithmic fit also should be expected for all reinforcement learning algorithms for practical reasons. Specifically, the reinforcement learning strategies, after receiving lots of data, should begin to reach a constant average payoff since adding more payoffs barely changes the weight distribution after many rounds.

Chord specific reinforcement consistently performed worse than simple reinforcement, which was true when comparing their performances against each other when playing any given available strategy. This was also true in Figure~\ref{fig4} indicating how each reinforcement learning strategy played against itself and where simple reinforcement learning quickly becomes better than chord specific reinforcement. This performance difference between simple and chord specific reinforcement learning was a surprising result. Music theory holds that certain notes harmonize better with different chords. Thus, a strategy that has the same probabilities regardless of the underlying chord would reasonably be expected to do worse than a strategy that learned different probabilities for each chord. It is possible that the Chord-Specific Reinforcement Learning strategy did worse because it simply lacked sufficient time to explore, adapt and learn. This slower learning rate can be attributed to its having a different but larger set of probabilities for each chord, resulting in an estimated learning rate that is about four times slower. This can be viewed in Figure~\ref{fig4}, as the blue dotted line representing the Chord-Specific Reinforcement Learning indeed has a slower ascent then the red dotted line representing simple reinforcement. In other words, a slower learning speed and fewer experiences led to consistently worse performance for Chord-Specific Reinforcement Learning strategy. However, if both curves continue at the same rate, Chord-Specific Reinforcement Learning would never catch up with Simple Reinforcement. At some point both curves will likely level off, but we currently have no direct evidence that Chord-Specific Reinforcement would level off higher. Another highly likely explanation is that the Chord-Specific Reinforcement can't properly adapt to the variance score. Because the simple reinforcement is always  updating the same probability table, it can quickly adapt to play less of a note if that note is played too often. However, the Chord-Specific strategy lacks this ability to react in a similar manner.

For example, imagine the following scenario: the Chord-Specific strategy develops a high probability for playing the note Bb (this is plausible because Bb harmonizes fairly well with nearly every chord). However, this note becomes played too much and then starts having a very negative variance score. The next time the Chord-Specific Reinforcement player plays the Bb, it gets a very low payoff and reduces the probability of playing that Bb. However, it only updated the probability of playing Bb for that chord. When, it gets to the next chord, it still has a high probability of playing Bb and makes the same mistake again. This repeats for every chord. The simple reinforcement strategy would not run into this issue. This example shows that the Chord-Specific strategy's poor performance could be due to slow reaction time to the variance score.

It is noteworthy that the highest standard deviation is associated with the Harmony Prediction strategy when played against itself. The standard deviation of this pair is 2.3 times the next largest standard deviation (Harmony Prediction vs. Stepwise Changes, see Table~\ref{tablesd}). The Harmony Prediction strategy's dependence on the starting note explains this high standard deviation. If both players start with notes that harmonize with each other, this will cause a feedback loop where each player will pick a note that is highly likely to harmonize again. However, if the note doesn't initially harmonize, each player will now be stuck continually picking notes that likely don't harmonize together.

\subsection{Limitations of the present approach}

Few papers have previously modeled music interactions using mathematics , especially in the context of game theory. Thus, this paper proposed a novel modeling framework to study jazz improvisation. While our game theoretical model represents an initial step in the direction of creating precise quantitative studies of jazz improvisation, it may still be considered in its early stages and has some limitations.

One primary issue stems from striving towards such a music model of payoffs itself: ``good'' music may never truly be able to be quantified. Humanity has varied tastes in music; what sounds good to one person does not always sound good to another. These variances can depends on many factors, including location and culture, gender, socio-economic factors, and age~\cite{liu2018relation, ferwerda2017personality}. Most concepts of variance and harmony, furthermore, arise from Western music~\cite{leung2011harmony}. Even basic principles of consonance and dissonance are not universal to humans, meaning that any quantitative musical models explored will likely have bias towards some culture ~\cite{mcdermott2016indifference}.

Another issue with this form of a reinforcement learning model are human limits. Human working memory precludes any realistic version of the model's reinforcement learning strategies over the entirety of the game. All of these algorithms rely on a weighting system based on all previous payoffs. By the end of this model, only a computer, not a human, can realistically remember and compute these weights while making decisions for this game. There is an argument to be made that humans learn their musical ``weights" over the course of years, and therefore the reinforcement learning models are simulating a sped-up version of this learning. However, future models may still benefit from establishing a cutoff (possibly gradual based on how long ago a note was played) of remembering notes and relevant payoffs. Such quantitative gradual memory-based models have been used to model harmonic tension via weighting functions and could be applied to a mathematical game~\cite{nikrang2018automatic}. It should also be taken into account that jazz musicians and improvisers tend to have stronger auditory working memory than non-musicians and even other musicians~\cite{johnson2002jazz, nichols2018score, hansen2013working}. Thus, determining such a gradual cutoff may prove difficult, and relevant literature on the working memory of jazz improvisers to make this determination may not be currently present at all.

Each component of the game theoretical model, the variance score and harmony score, is not a perfect representation of the real musical world. To maximize the variance score, a player would play every note (different octaves being treated the same) precisely the same number of times. While music demands variance to avoid feelings of boredom and repetition, this level of calculation in one's musical improvisation defeats the point of improvisation. Thus, the variance score incentivizes levels of calculation that may be contrary to the ideals of improvisation itself. 

Furthermore, the current calculation of the harmony score, which involves a strict-tolerance check of the greatest common divisor of the two notes, likely incentivizes low notes. In this strict-tolerance check, low notes are likelier to harmonize with other notes since pairs of lower numbers provide lower greatest common divisors. In several well-performing strategies (especially reinforcement learning strategies like Stepwise Changes and Two-Player Reinforcement Learning), we saw the frequency of the note played decrease over time. 

\subsection{Potential Future Research Prospects and Outlook}

The process of modeling music gives rise to several options. As with any mathematical model, simplifications must be made. In our case, we chose to only look at pitch with integer values and only used two scores for harmony and pitch variance. While this approach yields interesting results, future research could include models accounting for rhythm, phrasing, dynamics, and other factors. For example, a rhythm score could be designed by leveraging the current literature on mathematical analysis of rhythm. Each player would be given the option rest (play no notes), and then their overall rhythm could be analyzed in accordance with the Rhythmic Euclidean Algorithm~\cite{toussaint2005euclidean}. 

To add dynamics to the game, one could give both players the option to play notes at different volumes. One could then create two scores: a volume matching score, and a volume variance score. The first would reward players for playing a volume similar to their partner. The second would take the dynamic contrast throughout the song and reward players for playing many different volume levels (similar to the pitch variance score).

Further scores that may be added to a quantitative musical model, including in jazz improvisation, include scores for pitch, melody, timbre, and tonality of individual or small groups of notes at a time. These four features of music have been shown neuroscientifically to be distinct and thus would be relatively independent~\cite{alluri2012large}. Similarly, meter, tempo, and patterns of music as a whole activate different areas of the brain and could thus be individual scores~\cite{thaut2014human}. Combined with refined harmony and variance scores, precise quantitative music models are likely possible but knowing their correlation with human music taste would remain an unknown, perhaps unsolvable, issue.

This refinement of harmony and variance score calculations can happen through additional data analysis. For example, one could statistically measure pitch variance scores across a large data base of jazz solos such as the Weimar Jazz Database~\cite{Pfleiderer:2017:BOOK}. From this, one could create a probability distribution of achieving a particular variance score. With this probability distribution, one could create a more refined payoff calculation. Similar curves could be created for all other scores. 

The harmony score could also be refined in other ways, such as allowing a slight error in the frequencies to be calculated (for example, 4001 and 40 could still give a harmony score of 101). This change would account for human hearing not noticing slight differences in frequency. The minimum difference for change to be perceived at least half the time, known as the just-noticeable difference, is approximately 1\%~\cite{stolzenburg2015harmony}.

Additionally, the variance score could include different types of musical variance proposed for generative music previously. These include embellishment and melodic motion~\cite{berndt2012}. A variance score accounting for these types of variance could be more in-depth and may more closely approach the human notion of true musical variance.

The rules of the game can also be modified to involve other realistic scenarios. For example, as mentioned before, players could choose to skip particular beats. To enable this, the calculations of the harmony and variance scores would need to be modified to account for skips properly. However, this skipping of beats could play a role in the calculation of a potential rhythm score. Such a score could also be modeled of databases of present-day jazz music.
       
A network game can also be made, with a population of musicians with different strategies, and switch strategies of musicians that have low average payoffs after their interactions (2-player jazz improvisation games).

Future research may also involve machine learning. This includes several possible avenues. Artificial intelligence models could be used to analyze audio clips of music and judge them based on the mathematical scores like the harmony and variance scores we created for our game. One could use these scores to experimentally validate scores, create new scores, and refine existing scores, realizing the greater potential of human-AI collaboration~\cite{chen2023ensuring}. Another possible avenue could be training machine learning models on existing jazz solos, and them having them play the game. However, this may have the downside of creating only music similar to what has already been created rather than new jazz music. By leveraging the latest development techniques using diffusion models~\cite{huang2024symbolic}, the adversarial nature of our present work emphasizes and initiates will help mitigate these issues in algorithmic composition. Pattern recognition strategies using artificial intelligence could also be employed in the context of the game itself. For example, instead of having a simple reinforcement model, we could incorporate Markov chains to create an algorithm that considers each note within context~\cite{franz1998markov}. Another interesting avenue is to create a simulation in which an artificial intelligence plays against itself for many iterations of the game and creates its own optimal strategies. Then, the music it produces could be analyzed to see how well it corresponds with human-made improvisation.

Finally, this game only allows for integer frequencies to be played by players. This limit makes the note choices more fine-grained at higher frequencies. This is due to musical notes increasing in octaves on an exponential scale since a higher octave is equivalent to doubling the frequency of a note. Thus, decimal frequencies could be allowed; however, this would require modifications to the current calculation of the harmony score.

In sum, our work paves the initial way for exciting applications extending beyond jazz, such as employing artificial intelligence to analyze audio clips for enhancing musical reward systems, and utilizing machine learning to train models using existing jazz solos to improve strategic gameplay approaches within our modeling framework.

\section*{Acknowledgements}
We thank Caroline Hammond and Mark Lovett for their invaluable support during the initial phase of this work.

\section*{Code availability}
The code used for the analysis in this paper can be found at: \url{https://github.com/vedanttapiavala/frequencyGameTheory}.

\appendix

\section*{Appendix: Reinforcement Learning Generated Music Samples}
A folder containing music from a representative trial of each strategy pair has been created, and the QR code to this folder is available as Figure~\ref{s1}. To generate this music, the computer simulated the tenth trial of each strategy pair as normal between two players playing the selected strategy pair. Then, a MIDI file containing the note data (for each beat, both players' notes were played at the same time) was generated in the code.

After that, the MIDI file was imported into Logic Pro Version 10.8.1, and the eighth notes were quantized into a swing pattern, a common practice for jazz improvisers ~\cite{butterfield2011jazz}. Several accompanying tracks were manually added to mimic a jazz band accompaniment. These included horns, upright bass, and drums. The strategies produced by the game were set to the "Classical Electric Piano" sound. The resulting processed music is available in the folder.

To hear unprocessed music, the code's \textit{Main.java} file can be edited with a two-line change. The first player, represented by the Player variable p1, can be changed to the relevant player for the first strategy. The second player, represented by Player variable p2, can be similarly changed. After running the code, the relevant MIDI file will be generated, which can then be played.

\setcounter{figure}{0}
\renewcommand{\figurename}{Figure}
\renewcommand{\thefigure}{S\arabic{figure}}

 \begin{figure}[htbp]
  \centering
  \includegraphics[width=.25\textwidth]{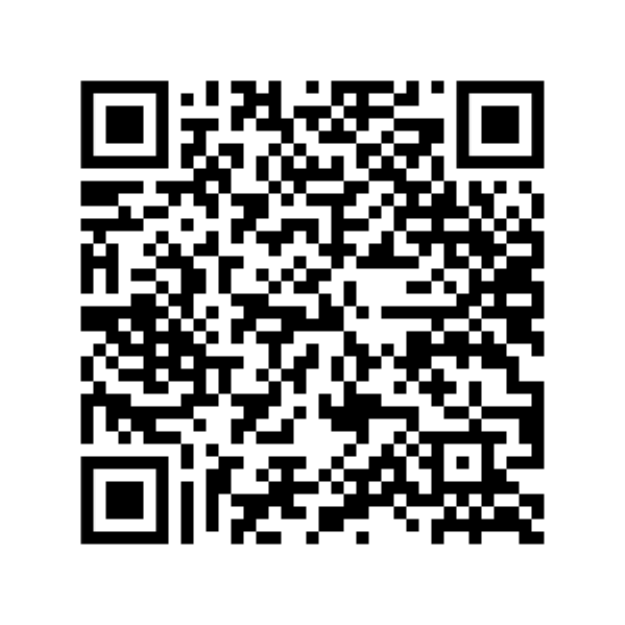}
  \caption{QR Code for Chosen Samples of Generated Music}
  \label{s1}
\end{figure}


\end{document}